\documentclass[nofootinbib,longbibliography,prd,twocolumn,showpacs,amsmath,amssymb]{revtex4-1}
\usepackage{dcolumn}
\usepackage{bm}
\usepackage{upgreek}
\usepackage{graphicx}
\usepackage{hyperref}

\newcommand{\rmd}{\mathrm{d}}
\newcommand{\rmi}{\mathrm{i}}

\newcommand{\W}{\varOmega}
\newcommand{\Wr}{\W_\text{r}}
\newcommand{\Wi}{\W_\text{i}}
\newcommand{\Wt}{\W_\text{t}}
\newcommand{\Wb}{\W_\text{b}}
\newcommand{\Wc}{\W_\text{c}}
\newcommand{\bpl}{\text{b}+}
\newcommand{\bmi}{\text{b}-}
\newcommand{\cpl}{\text{c}+}
\newcommand{\cmi}{\text{c}-}
\newcommand{\tpl}{\text{t}+}
\newcommand{\tmi}{\text{t}-}
\newcommand{\Kr}{K_\text{r}}
\newcommand{\Ki}{K_\text{i}}
\newcommand{\Kt}{K_\text{t}}
\newcommand{\Kb}{K_\text{b}}
\newcommand{\Kc}{K_\text{c}}
\newcommand{\D}{\mathfrak{D}}

\newcommand{\caP}{\mathcal{P}}

\newcommand{\tI}{\tilde{I}}

\newcommand{\V}{V}
\newcommand{\Vc}{V_\text{c}}
\newcommand{\N}{\mathfrak{n}}

\newcommand{\fK}{\mathfrak{K}}
\newcommand{\fW}{\mathfrak{K}^{-1}}
\newcommand{\caC}{\mathcal{C}}
\newcommand{\bbR}{\mathbb{R}}

\newcommand{\bfp}{\mathbf{p}}
\newcommand{\bfr}{\mathbf{r}}
\newcommand{\bfv}{\mathbf{v}}
\newcommand{\bfK}{\mathbf{K}}

\newcommand{\sfH}{\mathsf{H}}

\newcommand{\imag}{\text{Im}}

\newcommand{\diag}{\text{diag}}

\newcommand{\av}{\bar{v}}
\newcommand{\dv}{\sigma}

\newcommand{\AS}{\text{AS}}
\newcommand{\SB}{\text{SB}}

\newcommand{\tdf}[2]{\frac{\rmd #1}{\rmd #2}}
\newcommand{\tddf}[2]{\frac{\rmd^2 #1}{\rmd #2^2}}

\newcommand{\figscale}{0.72}

\begin{document}

\title{Dispersion relation of the fast neutrino oscillation wave} 

\date{\today}
\author{Changhao Yi}
\author{Lei Ma}
\author{Joshua D.\ Martin}
\author{Huaiyu Duan}
\email{duan@unm.edu}
\affiliation{Department of Physics \& Astronomy, University of New
  Mexico, Albuquerque, New Mexico 87131, USA}

\begin{abstract}
  A dense neutrino medium can support flavor oscillation waves which
  are coherent among different momentum modes of the neutrinos.
  The dispersion relation (DR) branches of such
  a wave with complex frequencies and/or wave numbers can lead to the
  exponential growth of the wave amplitude which in turn will engender
  a collective flavor transformation in the neutrino medium.
  In this work we propose that the complex DR branches of the neutrino
  oscillation wave should be bound by the critical points of the DR. We 
  demonstrate how this theory can be applied to the neutrino medium
  with an (approximate) axial symmetry about the propagation direction
  of the neutrino oscillation wave. We also show  how the
  flavor instabilities in this medium can be identified 
  by tracing the critical points of the DR as the
  electron lepton number distribution of the neutrino medium is
  changed continuously.  
\end{abstract}


\maketitle

\section{Introduction}

Through the neutrino-neutrino forward scattering
\cite{Fuller:1987aa,Notzold:1987ik,Pantaleone:1992xh}, the dense
neutrino gases present in the early universe, core-collapse 
supernovae and binary neutron star mergers can experience collective
flavor transformation (see, e.g.,
Refs.~\cite{Kostelecky:1993yt,Pastor:2002we,Duan:2006jv,Malkus:2012ts} among
many other works, and Ref.~\cite{Duan:2010bg} for a
review). Such a collective flavor transformation can play important
roles in the physical and chemical evolution of the environments of the
neutrino gases. This phenomenon becomes particularly interesting after
the realization that collective neutrino oscillations can occur at
very high densities and on
very tiny distance and time scales which are known as fast flavor
conversions \cite{Sawyer:2015dsa,
 Chakraborty:2016lct, Izaguirre:2016gsx,  Wu:2017qpc,
  Capozzi:2017gqd,
  Dasgupta:2016dbv, Abbar:2017pkh, Dasgupta:2017oko, Abbar:2018beu,
  Airen:2018nvp, 
  Abbar:2018shq} (see also Ref.~\cite{Chakraborty:2016yeg} for a review).

In the two-flavor-mixing scenario,
the flavor transformation of a neutrino can be viewed as the rotation
of the corresponding flavor (iso)spin $\vec{s}_\bfp$ in flavor
space \cite{Kim:1987bv,Duan:2005cp}, where $\bfp$ is the momentum of
the neutrino. It has been envisioned that the collective
transformation of the neutrinos is engendered by the ``neutrino flavor
spin wave''
\begin{align}
  \vec{s}_\bfp(t,\bfr)
  \propto e^{\rmi(\bfK\cdot\bfr-\W t)}
  \label{eq:spin-wave}
\end{align}
in the neutrino medium very much like the spin wave propagating
through a magnetic lattice, where $\bfK$ and $\W$
are the wave vector and frequency of the wave, respectively
\cite{Duan:2008v1}. However, most of the literature on collective
neutrino oscillations focused on the models of only one dimension
in either space or time. This paradigm was dramatically changed with
the recent discoveries of the spontaneous breaking of the spatial
and temporal symmetries by the collective neutrino oscillations
themselves (see, e.g.,
Refs.~\cite{Mangano:2014zda,Duan:2014gfa,Chakraborty:2015tfa,
  Mirizzi:2015fva,Abbar:2015fwa, Dasgupta:2015iia},
and also Ref.~\cite{Duan:2015cqa} for a review).
Since then, the dispersion relation (DR) approach and the 
instabilities of the DR have been introduced and investigated
\cite{Izaguirre:2016gsx,Capozzi:2017gqd,Wu:2017qpc,Airen:2018nvp}.

It is clear from Eq.~\eqref{eq:spin-wave} that a DR branch with a complex $\W$
and/or $\bfK$ 
can lead to an exponential growth of the amplitude of the neutrino
oscillation wave. The ranges of $\W$ and $\bfK$ where these complex DR
branches can exist, however, is not clear.
It was suggested in
Ref.~\cite{Izaguirre:2016gsx}  that the complex DR branches of the
neutrino oscillation wave 
always exist between the gaps of the real branches.
This conclusion turns out to be limited to 
the toy model with only two neutrino beams whose DR function is a quadratic
polynomial of the wave 
number and frequency \cite{Ma:2018}.
The introduction of the general theories and
classifications of the instabilities by Sturrock  and
Briggs in Ref.~\cite{Capozzi:2017gqd} was an important step forward,
which shows that some of the  
complex-$\bfK$ branches actually give rise the evanescent waves instead of
growing waves \cite{Sturrock:1958zz,Briggs:1964}.
Nevertheless, the actual calculations done in Ref.~\cite{Capozzi:2017gqd}
were still limited to the two-beam toy model. This is because the theories by
Sturrock and Briggs require the knowledge of the overall analytic properties
of the DR function which can be difficult to obtain for a general medium.
Partly due to this difficulty, it 
was later proposed in Ref.~\cite{Dasgupta:2018ulw} that the so-called ``zero
mode'', which supposedly represents the overall flavor evolution of
the neutrino medium,%
\footnote{It was disclosed in the discussion part of
  Ref.~\cite{Dasgupta:2018ulw} that it is possible for the system to
  be unstable in cases where the zero mode is stable. In other
  words, like any other Fourier mode, the zero mode may or may not be
  indicative of the overall flavor evolution of the medium.} 
be used to identify the potential fast
flavor conversions of the neutrinos in supernovae.

In this work we consider the
critical points of the DR function through the study of which some of the
insufficiency and misunderstandings of the previous works can be
addressed. The rest of the paper is organized as follows.
In Sec.~\ref{sec:DR} we identify the different types of
the critical points of the DR function, and we propose that the
complex DR branches should be bound by these critical points.
In Sec.~\ref{sec:axial-medium} we illustrate this theory by several
concrete examples which may be relevant to the fast flavor conversions
of the supernova neutrinos in the decoupling regime. Through these
examples We also demonstrate how
the flavor instabilities emerge and evolve as the neutrino angular
distributions of the medium changes.
In Sec.~\ref{sec:conclusions} we give our conclusions.

\section{Dispersion relation and instabilities}
\label{sec:DR}
\subsection{General theories}
Although the instabilities of the normal modes have been studied extensively
in other fields, it shall be helpful to recap some of the main
results here. We will mostly follow Sturrock's approach
\cite{Sturrock:1958zz} to the subject.
To demonstrate the basic concepts of
the instabilities, we
consider a one-dimensional medium with the wave equation
\begin{align}
  \D(\rmi\partial_t, -\rmi\partial_z) \psi(t,z) = 0,
  \label{eq:wave}
\end{align}
where $\psi(t,z)$ is the amplitude of the wave supported by the medium
at time $t$ and 
position $z$, and $\D(\rmi\partial_t,-\rmi\partial_z)$ is a linear
operator that contains the derivatives with respect to time and
space. A normal mode is a solution to
Eq.~\eqref{eq:wave} which has the  form
\begin{align}
  \psi(t,z) \sim e^{\rmi (K z - \W t)},
  \label{eq:NM}
\end{align}
where $K$ and $\W$ are constants. Such a
solution, if exists, implies the DR equation
\begin{align}
  \D(\W, K)=0.
  \label{eq:DR}
\end{align}
We will use $K=\fK(\W)$ and $\W=\fW(K)$ to denote the
DR in terms of $\W$ and $K$, respectively.

In general, both $\W$ and $K$ in Eq.~\eqref{eq:DR} can be complex:
\[ \W =\Wr + \rmi \Wi
\quad\text{and}\quad
K = \Kr + \rmi \Ki.\]
However, in practice one usually focuses on three kinds of DR branches:%
\footnote{We use the phrase ``branch'' to refer to a continuous,
  sometimes multi-valued, DR function $\fK(\W)$ for real
  $\W$ or $\fW(K)$  for real $K$.} 
\begin{itemize}
  \item the real branches along the real axis of $\W$ with
    $\fK(\W\in\bbR)$ also being real,
  \item the complex-$K$ branches along the real axis of $\W$ but with
    $\fK(\W\in\bbR)$ being complex,
  \item and the complex-$\W$ branches along the real axis of $K$ with
    $\fW(K\in\bbR)$ being complex.
\end{itemize}
A real DR branch gives rise to propagating waves in the medium with
$K$ and $\W$ being the wave number and the frequency of the wave, respectively.
The complex-$\W$ and complex-$K$ branches have been called the temporal and
spatial instabilities in the recent literature on neutrino oscillations because
of their apparent connections to the unstable waves and the amplifying waves
whose amplitudes grow in time and space, respectively. However, these
apparent connections are not always true. For example, it was
pointed out by Sturrock \cite{Sturrock:1958zz} that a complex-$K$ branch
without a companion complex-$\W$ branch gives rise to evanescent waves instead
of amplifying waves.

To illustrate Sturrock's theory of instabilities, we
 consider a spatially localized perturbation at $t=0$:
\begin{align}
  \psi(0,z) = \int_{-\infty}^\infty \zeta(K) e^{\rmi K z}\,
  \frac{\rmd
    K}{2\pi},
  \label{eq:psi0}
\end{align}
where $\zeta(K)$ is an analytic function of $K$ that peaks at $K_0$ and has
a finite spread.
If Eq.~\eqref{eq:wave} is a first-order differential equation in time,
and if there exists a DR branch $\fW(K\in\bbR)$, then 
\begin{align}
  \psi(t,z) =  \int_{-\infty}^\infty \zeta(K) \exp[\rmi K
    z-\rmi \fW(K) t]
  \,\frac{\rmd K}{2\pi}
  \label{eq:psit}
\end{align}
is the solution to Eq.~\eqref{eq:wave} that satisfies the initial
condition in Eq.~\eqref{eq:psi0}. Equation~\eqref{eq:psit} 
describes a ``space-like packet'' which is bound in space at any
given time. If $\fW(K)$ is a real branch, 
the initial wave packet is transported with the group velocity
\begin{align}
  V_0=\left(\tdf{\fW}{K}\right)_{K=K_0}.
\end{align}
If, however, (part of) $\fW(K)$ is a complex-$\W$ branch, and
$\imag[\fW(K)]$ is positive for a range of 
$K$ where $|\zeta(K)|$ is appreciable,
the amplitude of the wave in Eq.~\eqref{eq:psit} grows 
exponentially in time. This indicates the presence of the temporal
instabilities of the normal modes.

\begin{figure*}[htb]
  \begin{center}
    $\begin{array}{@{}c@{\hspace{0.05in}}c@{\hspace{0.05in}}c@{}}
      \includegraphics*[scale=\figscale]{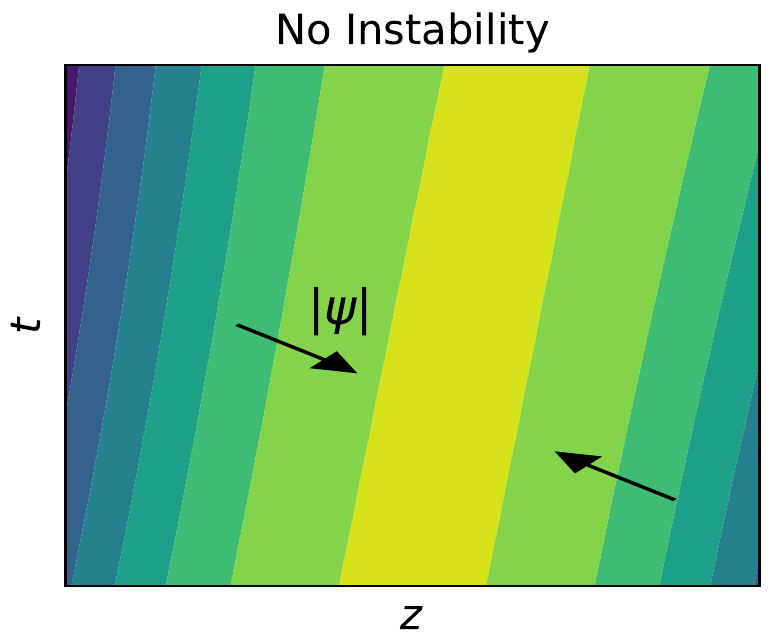} &
      \includegraphics*[scale=\figscale]{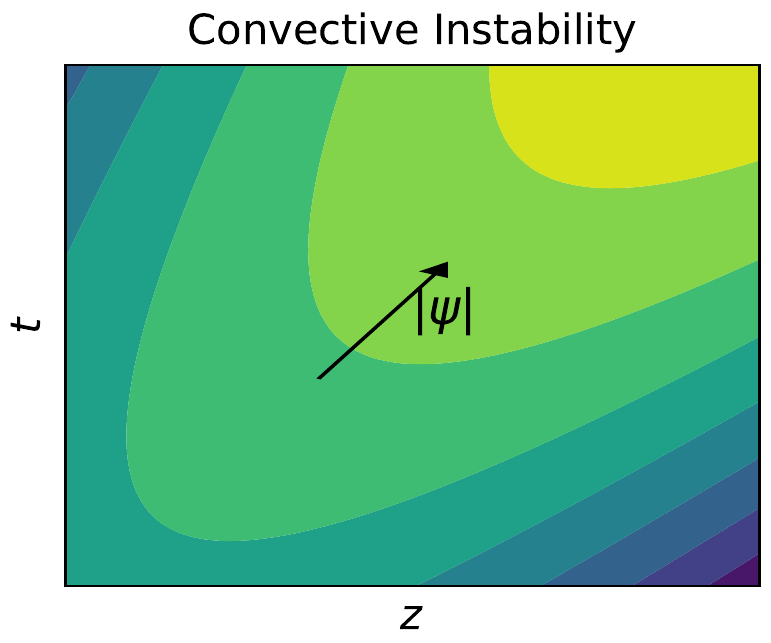} &
      \includegraphics*[scale=\figscale]{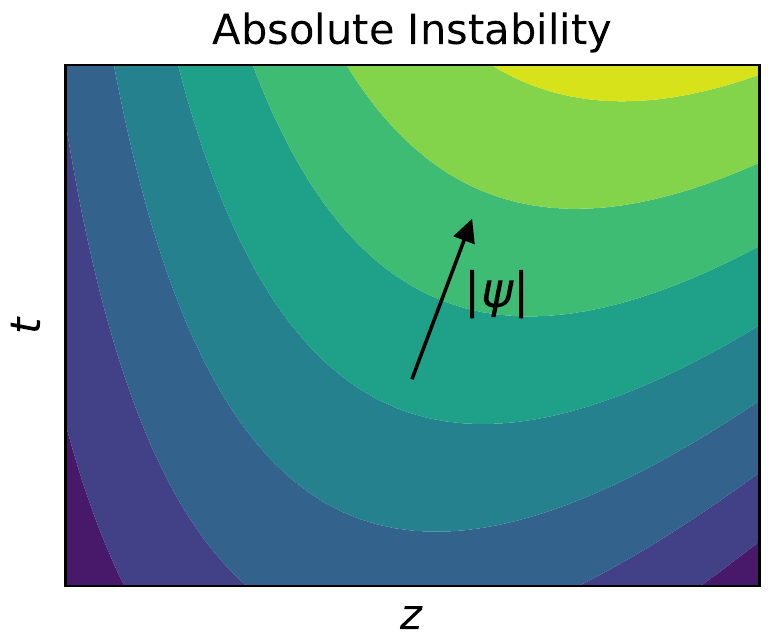} 
    \end{array}$
  \end{center}
  \caption{The evolution of three waves with no instability (left), a convective
    instability (middle), and an absolute instability (right),
    respectively. The arrow indicates the direction of increasing wave
    amplitude $|\psi|$.}
 \label{fig:psi}
\end{figure*}

A temporal instability can be either
convective or absolute (i.e.\ non-convective).
If the perturbation moves away from the point of
its origin as the wave amplitude grows, the corresponding instability is
convective. 
If, as the perturbation grows in both amplitude and
extent, it still embraces the point of the origin, the instability is
absolute. We illustrate the evolution of three waves with no
instability, with a convective instability, and with an absolute
instability in Fig.~\ref{fig:psi}.
Whether an instability is
convective or absolute depends on the reference frame of the
observer. Nevertheless, 
it is useful to make this distinction
because there usually is a reference frame that one prefers to work with.

Sturrock noted that, if the instability is convective,
$\psi(t,z)$ is not only bound
in $z$ at any given $t$ and but also bound in $t$ at any given
$z$. Therefore, at any spatial point,
$\psi(t,z)$ can be expressed in the form of a ``time-like packet'':
\begin{align}
  \psi(t,z) =  \int_{-\infty}^\infty
  \xi(\W) \exp[\rmi \fK(\W)
    z-\rmi \W t]
  \,\frac{\rmd \W}{2\pi},
  \label{eq:psit-W}
\end{align}
where  $\xi(\W)$ is an analytic function of $\W$
with a finite spread.  Meanwhile,
Eq.~\eqref{eq:psit} can also be rewritten as
\begin{align}
  \psi(t,z) =  \int_{\caC}
  \left[\zeta(\fK(\W)) \frac{\rmd\fK}{\rmd\W}\right]
  \exp[\rmi \fK(\W)
    z-\rmi \W t]
  \,\frac{\rmd \W}{2\pi},
  \label{eq:psit-K}
\end{align}
where $\caC$ is the path in the complex plane of $\W$ defined by
$\W=\fW(K\in\bbR)$. Sturrock concluded that, if the path $\caC$
could be continuously deformed in the complex plane of $\W$ to  the
real axis, the instability is convective, and
\begin{align}
  \xi(\W) = \zeta(\fK(\W)) \frac{\rmd\fK}{\rmd\W}.
\end{align}
Otherwise, the instability is absolute.

In Eq.~\eqref{eq:psit-W}, at least part of the DR branch
$\fK(\W\in\bbR)$ must be 
complex, or the wave will be stable. Therefore, a convective
instability implies at least one companion 
complex-$K$ branch to the complex-$\W$ branch, and the presence of
a complex-$\W$ branch alone gives rise to absolute
instabilities. However, we emphasize that the above reasoning does mean that the
instability is always convective when both both complex-$\W$ and
complex-$K$ branches are present.
As it will be shown later, an absolute instability can exist when the
complex-$\W$ and complex-$K$ branches do not form a closed contour on
the complex plane of $\W$ (with the help of the real branch).

Using the symmetry between $t$ and $z$ one can easily see that the
amplitude of the wave in a medium with 
a complex-$K$ DR branch alone will look similar to
Fig.~\ref{fig:psi}(c) except with 
$t\leftrightarrow z$. Therefore, such a complex-$K$ branch gives rise to
evanescent waves which die down as $t\rightarrow\infty$ at any spatial
point.

Sturrock's theory does not address the importance of the branch points
of the DR function $\fK(\W)$.
It was pointed out by Briggs \cite{Briggs:1964} that a branch point $(\Wb, \Kb)$
in the upper complex plane of 
$\W$ would result in absolute instabilities. In this case, the wave
amplitude at a given spatial point $z$ has the following asymptotic limit 
\begin{align}
  \psi(t, z) \propto \frac{e^{\rmi(\Kb
      z - \Wb t)}}{\sqrt{t}}
  \quad\text{ as }
  t\rightarrow \infty.
\end{align}

\subsection{The dispersion relation of the fast neutrino flavor conversion}

The DR for fast neutrino flavor conversions was first derived in
Ref.~\cite{Izaguirre:2016gsx}. Here we briefly review its main results 
for the convenience of the readers and also to
establish the formalism.
We consider the mixing between two neutrino flavors, $\nu_e$ and
$\nu_x$.
In a dense neutrino medium with all the
neutrinos initially in the weak-interaction states,
the flavor content of a
neutrino momentum mode $\bfp$ at time $t$ and location $\bfr$ can be
described by the flavor density matrix 
\cite{Sigl:1992fn,Izaguirre:2016gsx}
\begin{align}
  \rho_\bfp(t,\bfr) = \frac{f_{\nu_e} + f_{\nu_x}}{2}
  + \frac{f_{\nu_e} - f_{\nu_x}}{2} \begin{bmatrix}
    s & S \\ S^* & -s \end{bmatrix},
  \label{eq:rho}
\end{align}
where $f_{\nu_e/\nu_x}(\bfp)$ are the initial occupation numbers
of the corresponding neutrino flavors, the real field $s_\bfp(t,\bfr)$
and the complex field $S_\bfp(t,\bfr)$ describe the flavor conversion
and the flavor coherence of the neutrino, respectively.

In the absence of collisions, the neutrino flavor density matrix obeys
the equation of motion 
\cite{Sigl:1992fn}  
\begin{align}
\rmi (\partial_t + \mathbf{v} \cdot \bm{\nabla})
\rho_{\mathbf{p}} = \left[
  \frac{\mathsf{M}^2}{2\varepsilon} + \sfH_\text{mat} +
  \sfH_{\nu \nu,\bfp} ,
  \rho_{\mathbf{p}}\right],
\label{eq:eom}
\end{align}
where $\varepsilon=|\bfp|$, $\bfv=\bfp/\varepsilon$ and
$\mathsf{M}^2$ are the energy, velocity and the mass-square matrix of the
neutrino, respectively.
In the above equation,
\begin{align}
  \sfH_\text{mat} &= \sqrt{2} G_\text{F} n_e \begin{bmatrix}
    1 & 0 \\
    0 & 0 \end{bmatrix}
  \intertext{and}
  \sfH_{\nu\nu,\bfp} &= \sqrt{2} G_\text{F}
  \int\!(1-\bfv\cdot\bfv')(\rho_{\bfp'} - \bar\rho_{\bfp'})\,
  \frac{\rmd^3 p'}{(2\pi)^3}
\end{align}
are the matter and neutrino potentials, respectively, where
$G_\text{F}$ is the Fermi constant, $n_e$ is the electron number
density, and $\bar\rho_{\bfp}$ is the flavor density matrix of the
antineutrino which takes a form similar to Eq.~\eqref{eq:rho}.

We focus on fast neutrino flavor conversions which occur on very short
distance and time scales over which the physical conditions
such as $n_e$ and 
$f_\nu(\bfp)$ ($\nu=\nu_e,\bar\nu_e,\cdots$) are essentially
constant. The fast flavor evolution of the neutrino is energy
independent because the only energy-dependent term
$\mathsf{M}^2/2\varepsilon$  in Eq.~\eqref{eq:eom} is much smaller than the
rest of the Hamiltonian. (However, see
Ref.~\cite{Airen:2018nvp} for a scenario where the slow and fast
oscillations mix.)
To be self-consistent, we also assume that no significant
flavor conversion has occurred so that $|S_\bfv|\ll1$ and $s_\bfv\approx 1$.
In this case, it is useful to define the electron lepton number (ELN)
distribution 
of the neutrino \cite{Izaguirre:2016gsx}
\begin{align}
  G(\bfv) =\sqrt{2} G_\text{F}  \int_0^\infty
  [(f_{\nu_e} - f_{\bar\nu_e}) - (f_{\nu_x} - f_{\bar\nu_x})] \,
  \frac{\varepsilon^2 \rmd \varepsilon}{(2\pi)^3},
\end{align}
the ELN density
\begin{align}
  \varPhi_0 &= \int\!G(\bfv) \,\rmd\varGamma_\bfv,
\end{align}
and the ELN flux density
\begin{align}
  \bm{\Phi} &= \int\! G(\bfv)\bfv\,\rmd\varGamma_\bfv,
\end{align}
where $\rmd\varGamma_\bfv$ is the differential solid angle in the
direction of $\bfv$. Keeping only the terms of $\mathcal{O}(|S_\bfv|)$ or
larger in Eq.~\eqref{eq:eom}, one obtains
\cite{Banerjee:2011fj,Izaguirre:2016gsx} 
\begin{align}
  v^\beta[\rmi \partial_\beta 
   - (\varLambda_\beta + \varPhi_\beta)]S_\bfv
  = -v^\beta\int\! v'_\beta G(\bfv') S_{\bfv'}\rmd\varGamma_{\bfv'},
  \label{eq:eom-linear}
\end{align}
where $\mathrm{v}=[1,\bfv]$ is the four-velocity of the neutrino, and
$\Lambda = [\sqrt2G_\text{F} n_e, \mathbf{0}]$ and
$\Phi = [\varPhi_0, \bm{\Phi}]$ are the ELN fluxes carried by the
charged leptons and neutrinos, respectively.

For a normal mode of collective neutrino oscillations,
\begin{align}
  S_\bfv(t,\bfr) \sim e^{\rmi (\bfK\cdot\bfr - \W t)},
\end{align}
where $\W$ and  $\bfK$ are the frequency and wave vector of the
normal mode, respectively, both of which are independent of $\bfv$ or
the (initial) flavor of the neutrino. 
The mixing amplitude $S_\bfv$ of the normal mode 
grows exponentially in time if $\Wi>0$, and significant flavor conversions occur
when $|S_\bfv|\sim 1$. From Eq.~\eqref{eq:eom-linear} one sees
that it is convenient to make the following shifts to the frequency and
wave vector which do not affect the instabilities the normal modes:
\begin{align}
  \W - \varLambda_0 - \varPhi_0 \longrightarrow \W
  \quad\text{and}\quad
  \bfK -\bm{\Phi} -\bm{\Lambda} \longrightarrow \bfK.
  \label{eq:shift}
\end{align}
With these shifts, the DR for the fast flavor conversions in a
neutrino medium can be written as
\begin{align}
  \det[\Pi(\W, \bfK)] = 0,
  \label{eq:DR-gen}
\end{align}
where $\Pi$ is a $4\times4$ matrix with elements 
\begin{align}
  \varPi^{\beta\gamma} = \eta^{\beta\gamma}
  + \int \! G(\bfv)\,\frac{v^\beta v^\gamma}{\W - \bfK\cdot\bfv}\,\rmd
    \varGamma_\bfv
  \label{eq:Pi}
\end{align}
with $\upeta=\diag[+1,-1,-1,-1]$ being the metric tensor
of the Minkowski space. Because all the parameters and constants in
the above equation are real, if $(\W, \bfK)$ is a solution to
 Eq.~\eqref{eq:DR-gen}, so is $(\W^*, \bfK^*)$, where the
star indicates the complex conjugate.

\subsection{The limits of the complex branches}
\label{sec:crit-pts}
We assume that the
imaginary component of $\bfK$, if it is nonzero, is parallel or
anti-parallel to its real component which we
assume to be along the $z$ axis. With this assumption,
Eq.~\eqref{eq:Pi} can be written as
\begin{align}
  \varPi^{\beta\gamma} = \eta^{\beta\gamma}
  + \int_{-1}^1
  \frac{ G^{\beta\gamma}(v_z)}{\W - K v_z}\,\rmd v_z ,
  \label{eq:Pi-s}
\end{align}
where
\begin{align}
   G^{\beta\gamma}(v_z) = \int_0^{2\pi} G(\bfv) v^\beta
   v^\gamma\,\rmd\varphi
\end{align}
with
$\varphi$ being the azimuthal angle of $\bfv$ about the $z$ axis.
It has been noted \cite{Izaguirre:2016gsx} 
that a real branch does not exist in the 
``forbidden region'' where  $\varPi^{\beta\gamma}$ is
undefined.
In this region, the ``phase velocity''%
\footnote{Because of the redefinition of $\W$ and $\bfK$ in
  Eq.~\eqref{eq:shift}, $\V$ is not the actual phase velocity of the
  neutrino oscillation wave.}
of the normal mode
\begin{align}
 \V = \frac{\W}{K}
\end{align}
is within the range of $[-1,1]$.

In this subsection, we consider the limits of the complex branches.
We note that the DR function $K=\fK(\W)$ is (uniquely) 
defined by the DR equation $\D(\W,K)=0$ in the neighborhood of the
points where 
$\partial\D/\partial K$ is defined and is
nonzero. Therefore, we expect a complex-$K$ branch to end at
the critical points where $\partial\D/\partial K$ is either 0 or
undefined.%
\footnote{One may wonder whether a complex-$K$ branch can extend to
  $\W\rightarrow\pm\infty$. From Eq.~\eqref{eq:Pi-s} one sees that
  $\V$ of a complex-$K$ branch must approach a finite  
  real critical value $\Vc$ within the forbidden region as
  $\W\rightarrow\pm\infty$, or 
  $\varPi^{\beta\gamma}\rightarrow \eta^{\beta\gamma}$ and 
  Eq.~\eqref{eq:DR-gen} would not hold. Unless the
  ELN distribution is discontinuous, one can show that $\Vc$
  corresponds to a critical point of $\D(\W,K)$.}  
There are three types of
critical points for the complex-$K$ branches.
At a critical point of the first type,
$\partial\D/\partial K=0$. Near this point,
the DR equation can be written as \cite{Briggs:1964}
\begin{align}
  \left(\frac{\partial \D}{\partial \W}\right)_\text{b}(\W - \Wb)
+ \frac{1}{2}\left(\frac{\partial^2 \D}{\partial K^2}\right)_\text{b}
(K - \Kb)^2 \approx 0,
\label{eq:DR-b}
\end{align}
where we use the subscript `b' to indicates this type of critical
point. From the above equation one sees that $\Wb$ is a branch point
of $\fK(\W)$ on the complex 
plane of $\W$ around which $\fK(\W)$ is double valued.
Because $\Wb$ is real, $\Kb$ must also be real so that the branch
point is uniquely defined, which in turn implies that
$(\partial \D/\partial \W)_\text{b}$ and
$(\partial^2 \D/\partial K^2)_\text{b}$ are both real. From
Eq.~\eqref{eq:DR-b} one sees that
$(\Wb,\Kb)$ is a turning point of the real
branch $\W=\fW(K\in\bbR)$ where
\begin{align}
  \left.\tdf{\fW}{K}\right|_{K=\Kb} = 0,
  \label{eq:cond-b}
\end{align}
and where a conjugate pair of complex-$K$ branches connect to the
real branch.

A critical point of the second type is located at a
finite value of $\Wc$
where the phase velocity approaches a real nonzero value $\Vc$ within the
forbidden region. At this critical point $\partial\D/\partial K$ is
undefined. The two real values of $\Kc=\Wc/\Vc$ and $\Vc$ can 
be solved simultaneously from Eq.~\eqref{eq:DR-gen} by using 
\begin{align}
  K \varPi^{\beta\gamma} \xrightarrow{\V\rightarrow\Vc}
  \Kc\eta^{\beta\gamma} +
  \caP\!\int_{-1}^1
  \frac{ G^{\beta\gamma}(v_z)}{\Vc - v_z}\,\rmd v_z
  \pm \pi\rmi G^{\beta\gamma}(\Vc),
  \label{eq:Pi-c}
\end{align}
where the subscript `c' indicates the values at this critical point.
Here we have  used the Sokhotski-Plemelj theorem
\begin{align}
  \lim_{\epsilon\rightarrow0^+}\int_a^b
  \frac{f(x)}{x-\rmi\epsilon}\,\rmd x
  = \caP\!\int_a^b
  \frac{f(x)}{x-\rmi\epsilon}\,\rmd x
  + \pi\rmi f(0)
\end{align}
with the symbol $\caP$ denoting the principal value of the integral. 
The critical point $\Wc$ is also a branch 
point of $\fK(\W)$ on the complex plane of $\W$ where a conjugate pair
of complex-$K$ branches meet.
A special case of the critical points of the second type is where
$\Vc$ is a crossing point of the 
ELN distribution so that $G(\Vc, \varphi)=0$ for all $\varphi$.
In this case, $G^{\beta\gamma}(\Vc)=0$, and
\begin{align}
  K \varPi^{\beta\gamma} \xrightarrow{\V\rightarrow\Vc}
  \Kc \eta^{\beta\gamma}
  +\caP\!\int_{-1}^1
  \frac{ G^{\beta\gamma}(v_z)}{\Vc - v_z}\,\rmd v_z.
\end{align}
One can use the above expression to solve Eq.~\eqref{eq:DR-gen} for
$\Kc$ and $\Wc$.

The last type of the critical points is at $\W=0$ where
$\partial\D/\partial K$ is also undefined. Unlike a critical point of
the second type, $K_0=\fK(0)$ is complex which can be solved from
Eq.~\eqref{eq:DR-gen} by using 
\begin{align}
  K \varPi^{\beta\gamma} \xrightarrow{\V\rightarrow0}
  K_0\eta^{\beta\gamma}
  -\caP\!\int_{-1}^1
  \frac{ G^{\beta\gamma}(v_z)}{v_z}\,\rmd v_z
  \pm \pi\rmi G^{\beta\gamma}(0).
\end{align}

Similarly, a complex-$\W$ branch can end at two types of critical
points. At a type-I critical point  $(\Wt,\Kt)$, $\partial\D/\partial\W=0$. This
point is also a turning point of the real branch
$K=\fK(\W\in\bbR)$ with
\begin{align}
  \left. \tdf{\fK}{\W} \right|_{\W=\Wt} = 0,
  \label{eq:cond-t}
\end{align}
where we use the subscript `t' to indicate this type of critical points.
A conjugate pair of complex-$\W$ branches join a real branch at
$(\Wt, \Kt)$. 
A type-II critical point for the complex-$K$ branches is also a
type-II critical point for the complex-$\W$ branches where
both $\partial\D/\partial K$ and $\partial\D/\partial \W$ are undefined.
The points where $K=0$ are not critical because
$\partial\D/\partial\W$ is well defined there.

\section{Neutrino media with the axial symmetry}
\label{sec:axial-medium}

As a concrete example, we consider the neutrino media with an
(approximate) axial symmetry about the $z$ axis. 
For such a medium,  Eq.~\eqref{eq:DR-gen} gives two
DRs with different symmetry properties \cite{Izaguirre:2016gsx}:
\begin{align}
  \D_\AS(\W, K) = (I_0+1)(I_2-1) - I_1^2 = 0
  \label{eq:DR-AS}
\end{align}
and
\begin{align}
  \D_\SB(\W, K) = I_0 - I_2 - 2 = 0,
  \label{eq:DR-SB}
\end{align}
where
\begin{align}
I_k(\W, K) = \int_{-1}^1 G(v_z)\frac{v_z^k}{\W - K v_z}\,\rmd v_z
\end{align}
with
\begin{align}
  G(v_z) = \int_0^{2\pi} G(\bfv)\,\rmd\varphi.
\end{align}
Equation~\eqref{eq:DR-AS} is for the normal modes of axially
symmetric (AS) polarizations (i.e., $S_\bfv$ is independent of
the azimuthal angle $\varphi$), while
Eq.~\eqref{eq:DR-SB} is for those with
(axial-)symmetry-breaking (SB) polarizations
\cite{Raffelt:2013rqa,Izaguirre:2016gsx}.
The AS modes are the
mixtures of the monopole mode ($\ell=0$) and the axially symmetric
dipole mode ($\ell=1$ and $m=0$), and the SB modes
are  the linear superposition of the two degenerate dipole modes with $m=\pm1$
\cite{Duan:2013kba}.

\begin{figure}[htb]
  \begin{center}
    \includegraphics*[scale=\figscale]{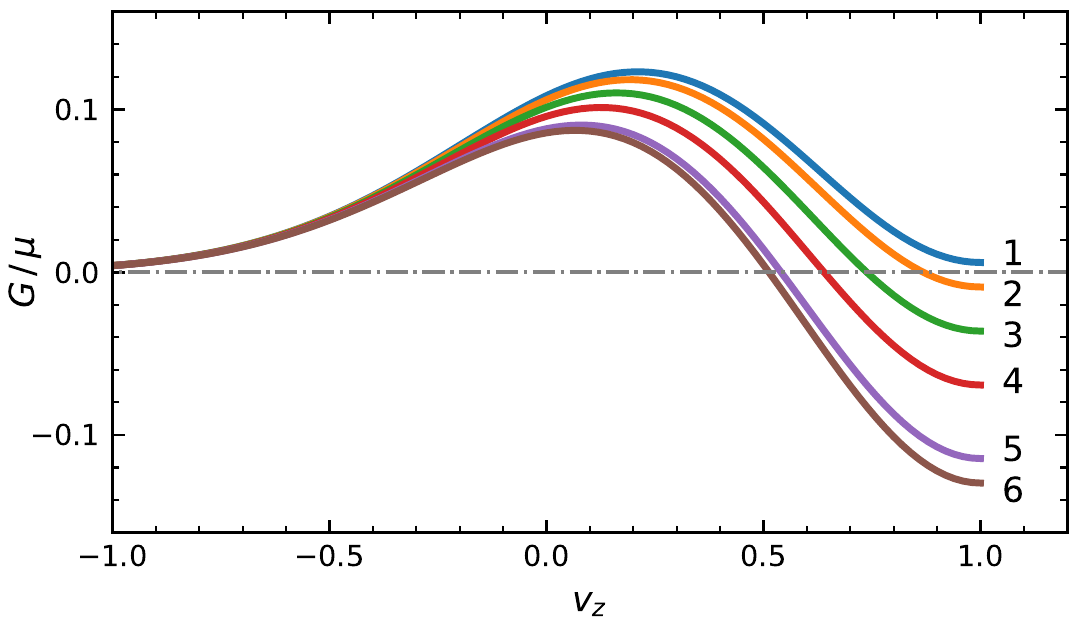}
  \end{center}
  \caption{The ELN distributions $G_i(v_z)$
    ($i=1,2,\cdots, 6$) used in the numerical examples as measured by
    the strength of the neutrino potential 
    $\mu=\sqrt2 G_\text{F} n_{\nu_e}$.}
 \label{fig:G}
\end{figure}

Because the typical distance scale of fast neutrino flavor conversions
is $\sim G_\text{F} n_\nu$, we define
\begin{align}
  \mu = \sqrt2 G_\text{F} n_{\nu_e}
\end{align}
to be the unit of all quantities with dimensions, where
\begin{align}
  n_{\nu_e} = \int\! f_{\nu_e}(\bfp)\,\frac{\rmd^3 p}{(2\pi)^3}
\end{align}
is the (initial) number density of $\nu_e$'s.
To demonstrate the typical features of the DRs of fast neutrino flavor
conversions, we employ a set of 6
parametrized distributions all of which are of the form
\begin{align}
  G_i(v_z) = \mu[g(v_z, \av_0, \dv_0) - \alpha_i g(v_z, \av_i, \dv_i)],
  \label{eq:Gi}
\end{align}
where $i=1,2,\cdots,6$,
\begin{align}
  g(v_z, \av, \dv) \propto \exp\left[-\frac{(v_z -
      \av)^2}{2\dv^2}\right]
\end{align}
is the Gaussian distribution with the normalization condition
$\int_{-1}^1 g\,\rmd v_z=1$, and the  values of $\alpha_i$, $\av_i$,
and $\dv_i$ are listed in Table~\ref{tab:G-par}.
All the distributions have positive ELN densities $\varPhi_0 >0$,
but their ELN flux densities $\varPhi_z$ in the $z$ direction
can be either positive or negative:
\begin{itemize}
\item $G_1$ is a distribution that 
  stays positive for all $v_z$;
  
\item $G_2$ has a ``shallow'' crossing and a positive
  $\varPhi_z$;

\item $G_3$ and $G_4$ have ``moderate'' crossings and negative
  $\varPhi_z$;

\item $G_5$ and $G_6$ have ``deep'' crossings and negative
  $\varPhi_z$.
\end{itemize}
These ELN distributions are plotted in Fig.~\ref{fig:G}. 

\begin{table}[htb]
\caption{\label{tab:G-par} The parameters used in the ELN
  distributions in Eq.~\eqref{eq:Gi}.} 
\begin{ruledtabular}
  \begin{tabular}{@{\hspace{0.2in}}cddd@{\hspace{0.2in}}}
    $i$ &
    \multicolumn{1}{c}{$\alpha_i$} &
    \multicolumn{1}{c}{$\av_i$} &
     \multicolumn{1}{c}{$\dv_i$} \\ \hline
  0 & \multicolumn{1}{c}{--} & 1.0 & 0.6 \\
  1 & 0.88 & 1.0 & 0.53 \\
  2 & 0.89 & 1.0 & 0.53 \\
  3 & 0.908 & 1.0 & 0.53 \\
  4 & 0.93 & 1.0 & 0.53 \\
  5 & 0.96 & 1.0 & 0.53 \\
  6 & 0.97 & 1.0 & 0.53 
\end{tabular}
\end{ruledtabular}
\end{table}

\subsection{Symmetry-breaking modes}

\begin{figure*}[htb]
  \begin{center}
    $\begin{array}{@{}c@{\hspace{0.01in}}c@{\hspace{0.01in}}c@{}}
      \includegraphics*[scale=\figscale]{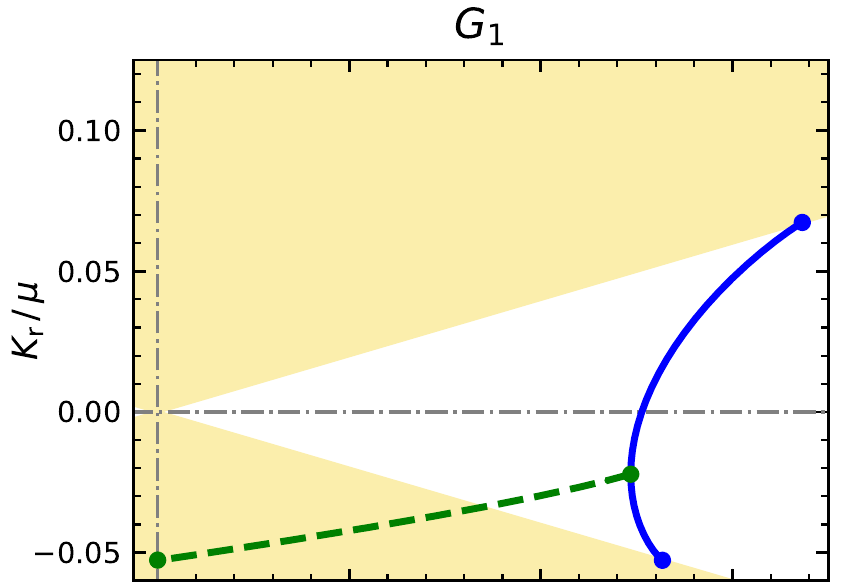} &
      \includegraphics*[scale=\figscale]{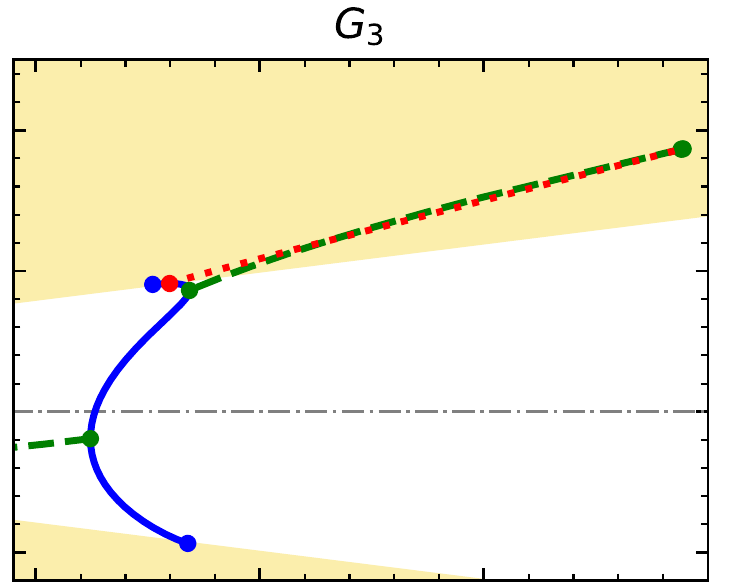} &
      \includegraphics*[scale=\figscale]{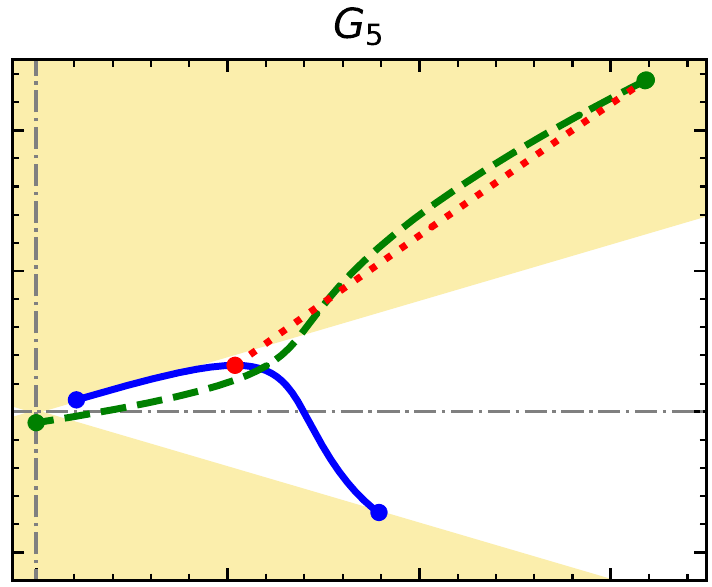} \\
      \includegraphics*[scale=\figscale]{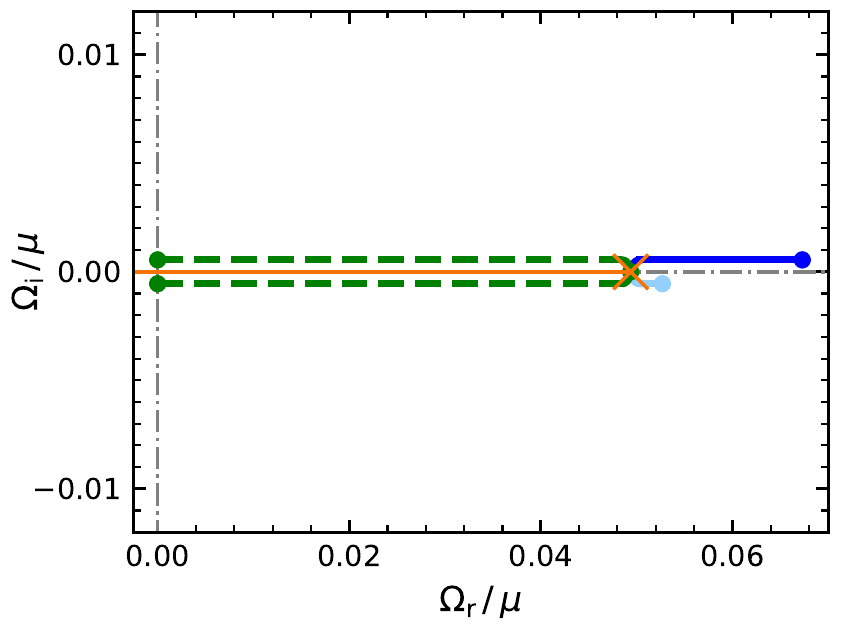} &
      \includegraphics*[scale=\figscale]{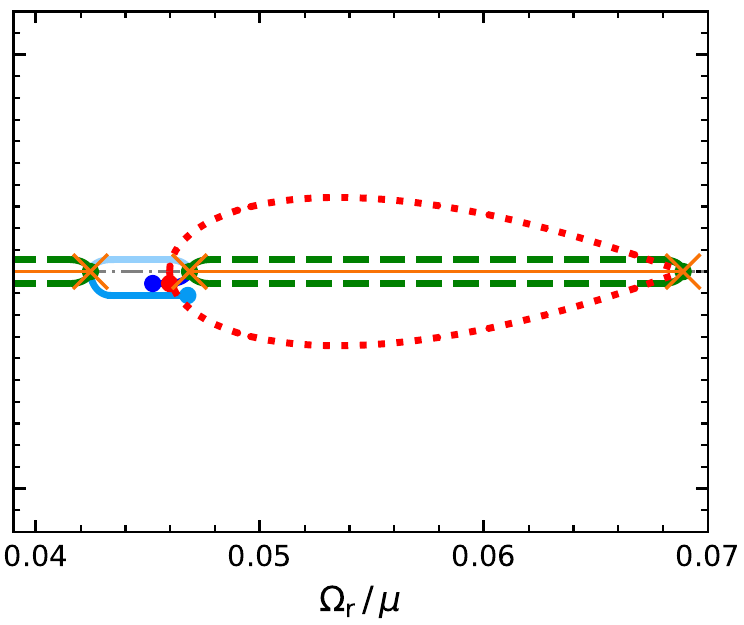} &
      \includegraphics*[scale=\figscale]{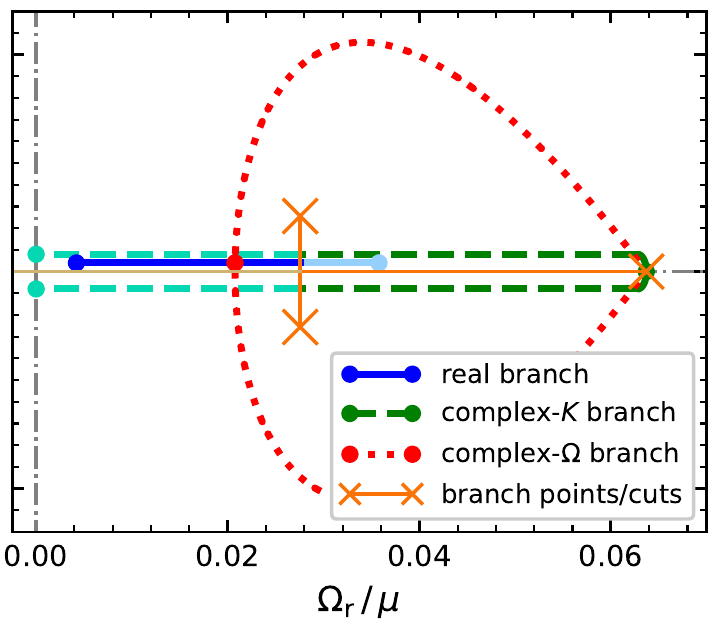} 
    \end{array}$
  \end{center}
  \caption{(Color online) The real (solid lines), complex-$K$ (dashed lines) and
    complex-$\W$ (dotted lines) DR branches for the SB modes with
    $G_1$ (left panels), $G_3$ (middle panels) and $G_5$ (right panels)
    distributions on the 
    $\Wr$-$\Kr$ plane (upper panels) and the complex plane of $\W$
    (lower panels), respectively. The filled circles represent the end
    points of the branches, and the crosses denote the branch points on
    the complex plane of $\W$ where two Riemann sheets join.
    The shadowed regions in the upper panels are the forbidden regions
    for the real branches.
    In the lower panels, the lines of the same style but
    different colors/intensities 
    represent the same DR branches on
    different Riemann sheets, and the DR branches
    along the real axis of the complex plane of $\W$ are shifted up or
    down for clarity. }
 \label{fig:DR-SB}
\end{figure*}

\subsubsection*{Real branches}

The real DR branches of the SB modes can
be readily solved from Eq.~\eqref{eq:DR-SB} in terms of the
``refractive index'' \cite{Izaguirre:2016gsx}
\begin{align}
  \N = \frac{1}{V} = \frac{K}{\W}.
\end{align}
For each value of $\N\in[-1,1]$ one obtains
\begin{subequations}
\label{eq:WK-SB}
\begin{align}
  \W(\N) &= \frac{1}{2}\int_{-1}^1 G(v_z)\,\frac{1- v_z^2}{1-\N v_z}\,\rmd v_z
  \label{eq:W-SB}\\
  \intertext{and}
  K(\N) &= \N \W.
\end{align}
\end{subequations}
As noted in Ref.~\cite{Izaguirre:2016gsx}, the real branches of the SB
modes always end at finite values of $\W$ and $K$ because
\begin{align}
\W(\N=\pm1) =  \frac{1}{2}\int_{-1}^1 G(v_z)(1\pm v_z)\,\rmd
v_z
\end{align}
are finite.

As mentioned in Sec.~\ref{sec:crit-pts}, the turning points of the
real branches are also critical points of the DR. To locate these
turning points, we differentiate
Eq.~\eqref{eq:WK-SB} with respect to $\N$ and obtain
\begin{subequations}
  \label{eq:WKn-SB}
\begin{align}
  \tdf{\W}{\N} &= \frac{1}{2}\int_{-1}^1 G(v_z)\,\frac{v_z(1- v_z^2)}{(1-\N
    v_z)^2} \,\rmd v_z,
  \label{eq:Wn-SB}\\
  \tddf{\W}{\N} &= \int_{-1}^1 G(v_z)\,\frac{v_z^2(1- v_z^2)}{(1-\N
    v_z)^3} \,\rmd v_z,
  \label{eq:Wnn-SB}\\
  \tdf{K}{\N} &= \frac{1}{2}\int_{-1}^1 G(v_z)\,\frac{1- v_z^2}{(1-\N
    v_z)^2}\,\rmd v_z.
  \label{eq:Kn-SB}
\end{align}
\end{subequations}
Based on the overall geometric shapes of their real DR branches, the 6
parametrized ELN distributions fall into three categories.

The first category of the ELN distributions includes $G_1$ which 
remains positive for the whole range of $v_z$. The real branch of an ELN
distribution in this category has the following properties:
\begin{itemize}
  \item $K(\N)$ is a monotonically increasing function of $\N$ because
    $\rmd K/\rmd \N>0$.
  \item Both $\W(\N)$ and $\rmd^2\W/\rmd \N^2$ stay positive for the
    whole range of $v_z$, but
    $\rmd\W/\rmd \N$ changes sign between $v_z=-1$ and $1$ because
\begin{align}
  \tdf{\W}{\N} &\xrightarrow{\N\rightarrow-1^+}
  G(-1) \ln(1+\N) <0\\
  \intertext{and}
  \tdf{\W}{\N} &\xrightarrow{\N\rightarrow 1^-}
  -G(1) \ln(1-\N) > 0.
  \label{eq:dWdn-1}
\end{align}
    This implies the existence of a turning point
    $(\Wb, \Kb)$ on the real branch where $\rmd\W/\rmd K=0$.     
\end{itemize}

The second category includes $G_2$ and $G_3$ which are slightly
negative in the forward direction ($v_z=1$). The real branch of a
category-II distribution is similar to that of category I but with
the following important differences:
\begin{itemize}
\item $\rmd K/\rmd \N$ changes sign near $\N=1$ because
  \begin{align}
    \tdf{K}{\N} &\xrightarrow{\N\rightarrow 1^-}
    -G(1) \ln(1-\N) < 0.
  \end{align}
  This implies the existence of a turning point $(\Wt, \Kt)$ on the
  real branch where $\rmd K/\rmd\W=0$.
  
\item $\rmd \W/\rmd \N$ changes sign again near $\N=1$ where
  $\rmd\W/\rmd\N$ becomes negative [see Eq.~\eqref{eq:dWdn-1}].
  This indicates 
  the appearance of another turning point $(\Wb', \Kb')$ on the
  real branch.
\end{itemize}

The last
category of the ELN distributions include $G_4$ through $G_6$ whose
real DR branches have only one critical point $(\Wt, \Kt)$.

We calculated the real DR branches for $G_1$, $G_3$
and $G_5$, which represent the three categories of ELN distributions,
and show them as solid curves in Fig.~\ref{fig:DR-SB}.

\subsubsection*{Complex-$K$ and complex-$\W$ branches}
\label{sec:complex-K}

For the SB modes, a type-II critical point $(\Wc, \Kc)$ can exist only
if the ELN distribution $G(v_z)$ has a crossing point $\Vc$. From the
above discussion one 
sees that, for a category-I  distribution, there are only two
critical points for the complex-$K$ branches,  $\W=0$ and $\Wb$.
Therefore, a pair of complex-$K$
branches run from $(\Wb, \Kb)$ to $(0, K_0)$ and $(0, K_0^*)$, respectively.
For a category-II distribution, there are two additional critical
points, $\Wb'$, and $\Wc$. As a result, another pair of complex-$K$
branches run from $(\Wc, \Kc)$ to $(\Wb', \Kb')$. For a category-III
distribution, there are again only two critical points, $0$ and
$\Wc$, and a pair of complex-$K$ branches run from $(\Wc, \Kc)$
to $(0, K_0)$ and $(0, K_0^*)$, respectively.

A category-I ELN distribution does not have any complex-$\W$ branch
because there does not exist any critical point for these branches. Both
category-II and category-III distributions have two critical points
which are located at
$K=\Kt$ and $\Kc$, respectively. Therefore, a pair of complex-$\W$
branches run from $(\Wc, \Kc)$ to $(\Wt, \Kt)$.

We calculated the complex-$K$ and complex-$\W$ branches for
$G_1$, $G_3$ and $G_5$, respectively, and show them as dashed and
dotted curves in Fig.~\ref{fig:DR-SB}.

\subsection{Identifying the instabilities}

The above results clearly show some of the misunderstandings in
the literature. For example, instead of being confined to the ``gap''
of the real branches as suggested by Ref.~\cite{Izaguirre:2016gsx},
the complex-$K$ branches for the $G_5$
distribution co-exist with the real branch in some range of $\Wr$. One
also sees that the complex-$\W$ branches exist for this distribution even
though the zero mode is stable, i.e.\ $\fW(K=0)$ is real.

The study of the DR branches of the SB modes reveals an interesting
pattern of how the instabilities appear and evolve as the ELN
distribution is changed continuously. One starts from a
category-I distribution without crossing which has no complex-$\W$
branch. For this 
distribution all the branch points $(\Wb, \Kb)$ are located on the
real axis of the complex plane of $\W$, and they are the turning
points on the real branch. There is no branch
point on the upper or lower complex plane which would imply the
existence of complex-$\W$ branches \cite{Briggs:1964}.

The instabilities may begin to occur as one varies the ELN
distribution. Because the strength of the instability increases 
with the magnitude of $\imag(\Wb)$, we expect that all the branch points
should appear on the real axis first before moving to the upper and
lower complex plane of $\W$.  As the ELN distribution begins to
develop a shallow crossing and become category II, two new turning points,
$(\Wb', \Kb')$ and $(\Wt, \Kt)$ appear on the reach branch.
The complex-$\W$ branch (with $\Wi>0$)  connecting to $(\Wt, \Kt)$
gives rise to convective instabilities because there is no branch
point on the upper complex plane of $\W$. This also agrees with
Sturrock's theory because the path $\caC$ of this complex-$\W$ branch can be
deformed continuously in the complex plane of $\W$ to a path along the
real axis which is made of a 
complex-$K$ branch and a segment of the real branch. (See the lower middle
panel of Fig.~\ref{fig:DR-SB}).

As the crossing of a category-II distribution becomes deeper and deeper, the
two turning points $(\Wb, \Kb)$ and $(\Wb', \Kb')$ on the real branch
come closer and closer to each other and finally merge into a saddle
point. Correspondingly, the two pairs of complex-$K$ branches merge
into a single pair. As the
crossing of the ELN distribution further deepens, the two branch
points of $\fK(\W)$ move to the upper and lower complex planes with
$\Wb'=\Wb^*$ and $\Kb'=\Kb^*$.  Although these branch points are not
on the real branch, their existence can be inferred by the fact that the
complex-$K$ and real branches pass each other without
intersection. (See the upper right panel of Fig.~\ref{fig:DR-SB}.) 
According to Briggs' theory, the instabilities
associated with the complex-$\W$ branch of a category-III 
distribution are absolute. This conclusion is also
in agreement with Sturrock's theory because the path of the
complex-$\W$ branch in 
the upper complex plane of $\W$ cannot be deformed to the real axis
due to the existence of the branch point.
(See the lower right panel of Fig.~\ref{fig:DR-SB}.)
This result can also be
deduced from the paths of the DR branches in the $\Wr$-$\Kr$ plane
where no closed loop is formed by the complex-$\W$ branch and other DR
branches with real $\W$.

\subsection{Axially symmetric modes}

\begin{figure*}[htb]
  \begin{center}
    $\begin{array}{@{}c@{\hspace{0.01in}}c@{\hspace{0.01in}}c@{}}
      \includegraphics*[scale=\figscale]{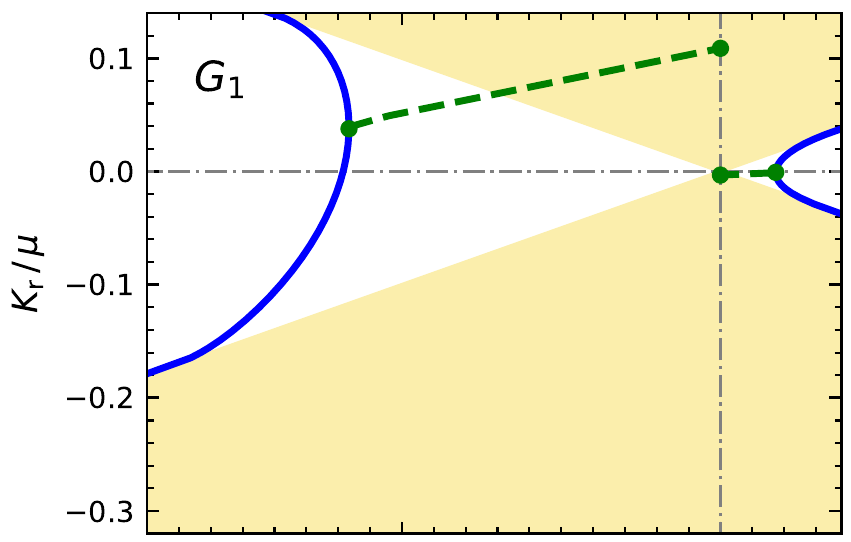} &
      \includegraphics*[scale=\figscale]{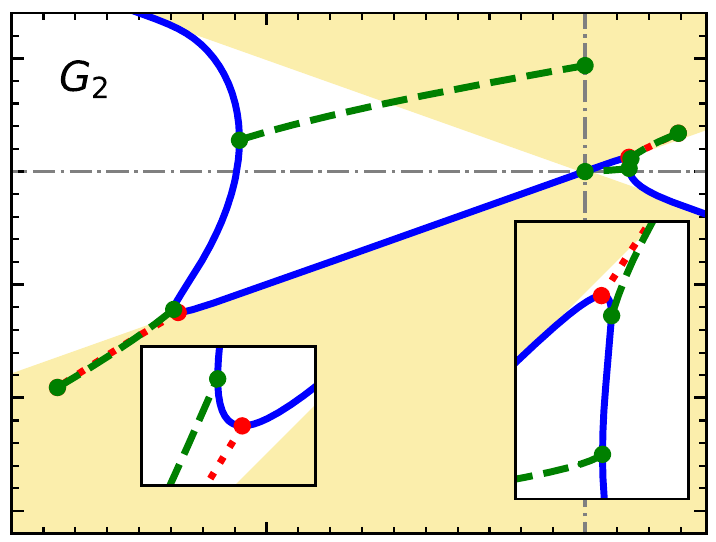} &
      \includegraphics*[scale=\figscale]{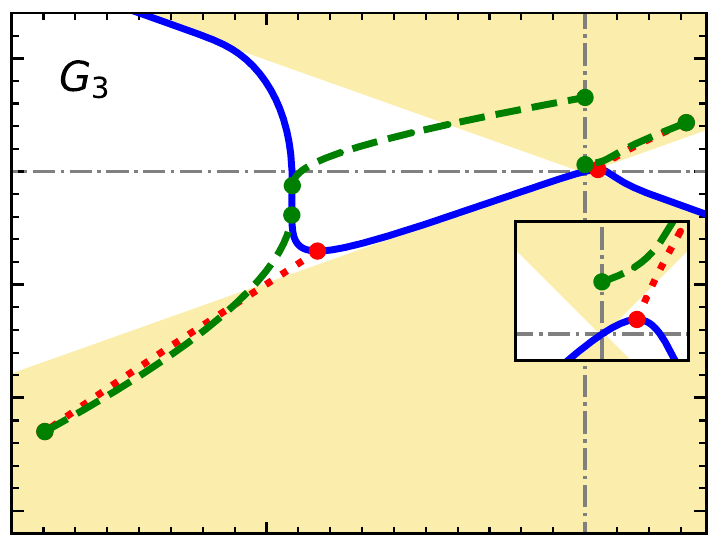} \\
      \includegraphics*[scale=\figscale]{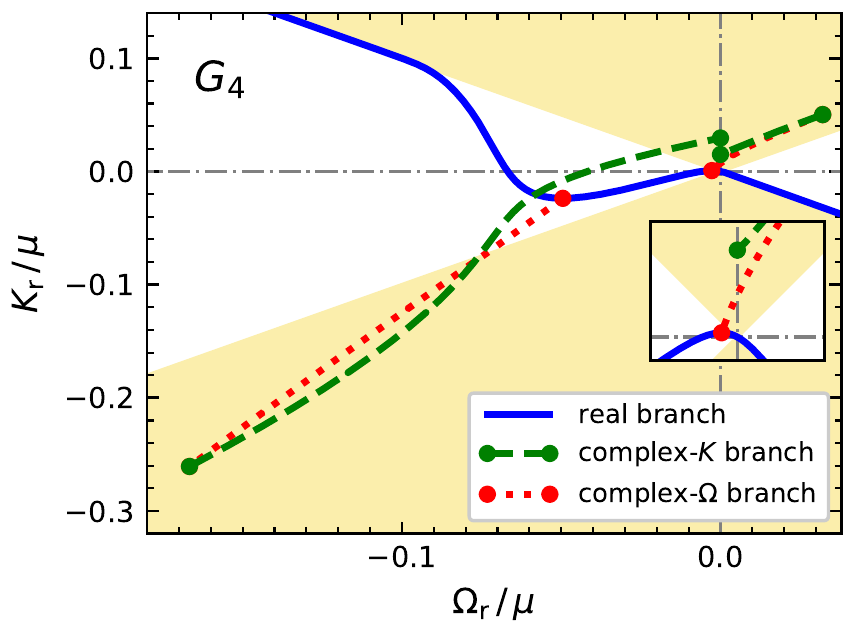} &
      \includegraphics*[scale=\figscale]{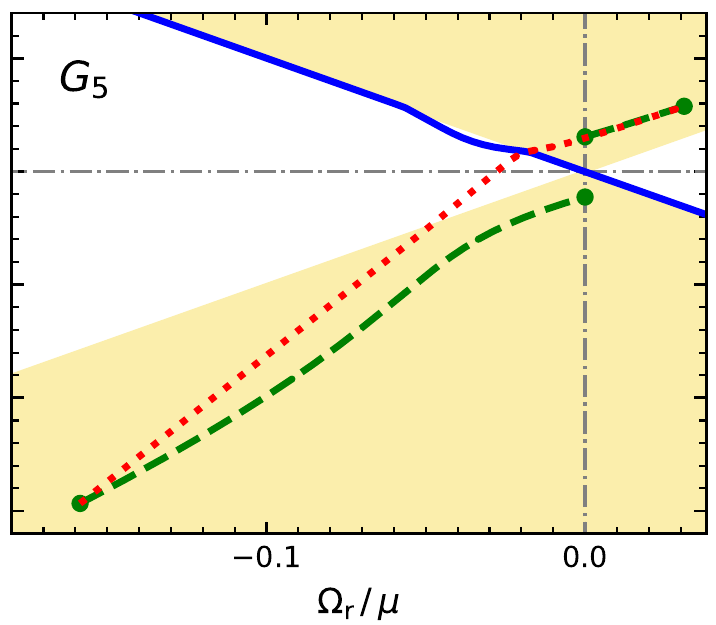} &
      \includegraphics*[scale=\figscale]{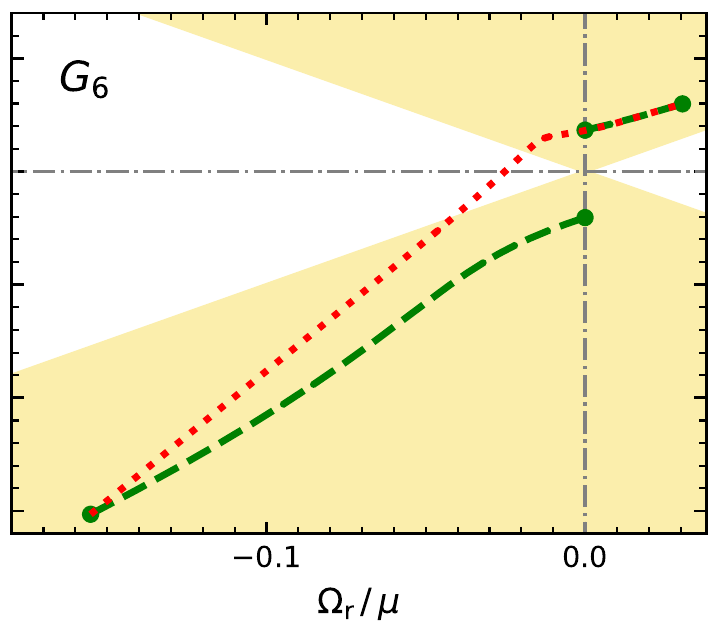} 
    \end{array}$
  \end{center}
  \caption{(Color online) Similar to Fig.~\ref{fig:DR-SB} but for the
    AS modes  with the ELN distributions $G_1$ through $G_6$. The
  small insets show the enlarged region around some of the critical
  points on the real DR branches.}
 \label{fig:DR-AS}
\end{figure*}

The real branches of the AS modes can also be
solved in terms of $\N$:
\begin{align}
  \W_\pm(\N) = \frac{\tI_2-\tI_0 \pm \sqrt\Delta}{2},
\end{align}
where
\begin{align}
  \tI_k(\N) = \int_{-1}^1 G(v_z) \frac{v_z^k}{1-\N v_z}\,\rmd v_z,
\end{align}
and
\begin{align}
  \Delta(\N) &= (\tI_2 - \tI_0)^2 + 4(\tI_2\tI_0 - \tI_1^2)
  \nonumber\\
  &= (\tI_0 + \tI_2 - 2\tI_1)(\tI_0 + \tI_2 + 2\tI_1).
\end{align}
The  evolution of the DR branches of the AS modes with the changing
ELN distribution is similar to that
of the SB modes but with a few new twists.
For the AS modes, the distributions $G_1$ through $G_6$ each has its
unique features and 
represents its own category. We calculated
the DR branches of the AS modes for all 6 distributions and plot them
in Fig.~\ref{fig:DR-AS}. 

\subsubsection*{Category I}

The distributions represented by $G_1$ has no crossing. For such a
distribution, $\tI_0>\tI_2>0$ and $\tI_2\tI_0>\tI_1^2$. Therefore,
$\W_\pm(\N)$ represent two separate real branches. In the limit
$\N\rightarrow\pm1$, $\W_\pm\rightarrow\pm\infty$ because
\begin{align}
  \Delta \xrightarrow{\N\rightarrow\pm1}
  -4G(\pm1)\ln(1\mp\N)\int_{-1}^1(1\mp v_z)G(v_z)\,\rmd v_z.
  \label{eq:Delta-lim}
\end{align}
The ``plus'' and ``minus'' real branches each has a critical point,
$(\W_{\bpl}, K_{\bpl})$ and $(\W_{\bmi}, K_{\bmi})$, and two conjugate
pairs of complex-$K$ branches run from these critical points to
$(0, K_{0+})$, $(0, K_{0+}^*)$, $(0, K_{0-})$, and $(0, K_{0-}^*)$,
respectively. There is no complex-$\W$ branch for a category-I
distribution because there 
exists no critical point associated with these branches.

\subsubsection*{Category II}
The distributions represented by $G_2$ has a shallow crossing near the
forward direction. Because $G(1)$ is slightly negative,
$\Delta\xrightarrow{\N\rightarrow1^-}-\infty$. As a result, the real
branch $\W_\pm(\N)$
is defined for $\N\in(-1, \N_\star]$, where
$\Delta(\N_\star)=0$. The originally separated two real branches bend
sharply toward each other near $V=1$ and merge into a single real
branch at $V=1/\N_\star$. Around 
the sharp bends of the real branch new critical points
$(\W_{\text{b}\pm}', K_{\text{b}\pm}')$ and
$(\W_{\text{t}\pm}, K_{\text{t}\pm})$ appear. Another pair of
critical points $(\W_{\text{c}\pm}, K_{\text{c}\pm})$ also
appear on the positive and negative sides of $\Wr$ both of which correspond to
the crossing point $\Vc$ of the ELN distribution. With the appearance
of the new critical points, two additional conjugate pairs of
complex-$K$ branches run from 
$(\W_{\text{b}\pm}', K_{\text{b}\pm}')$ to
$(\W_{\text{c}\pm}, K_{\text{c}\pm})$, and two conjugate pairs of
complex-$\W$ branches run from $(\W_{\text{t}\pm}, K_{\text{t}\pm})$ to
$(\W_{\text{c}\pm}, K_{\text{c}\pm})$. Both complex-$\W$ branches with
$\Wi>0$ give rise to convective instabilities.

\subsubsection*{Category III}
The distributions represented by $G_3$ have deeper crossings than
$G_2$. On the positive side of $\Wr$, the two branch points
$(\W_{\bpl}, K_{\bpl})$ and 
$(\W_{\bpl}', K_{\bpl}')$ have moved off the real branch to the upper
and lower complex planes of $\W$, and the two pairs of complex-$K$
branches have merged into a single pair. Correspondingly, the
instabilities associated with the
complex-$\W$ branch connecting to $(\W_{\cpl}, K_{\cpl})$ become absolute.

\subsubsection*{Category IV}
The distributions represented by $G_4$ have even deeper crossings than
$G_3$. At this point, the two branch points
$(\W_{\bmi}, K_{\bmi})$ and 
$(\W_{\bmi}', K_{\bmi}')$ have also moved off the real branch to the upper
and lower complex planes of $\W$, and the two pair of complex-$K$
branches on the negative side of $\Wr$ have merged into a single
pair. The complex-$\W$ branches on both the positive and
negative sides of $\Wr$ give rise to absolute instabilities.

\subsubsection*{Category V}
As the crossing of the ELN distribution becomes deeper and deeper, the
two turning points $(\W_{\tpl}, K_{\tpl})$ and $(\W_{\tmi}, K_{\tmi})$
move closer and closer to each other and eventually merge into a
saddle point before disappearing from the real branch. For the $G_5$
distribution, the two pairs of complex-$\W$ branches have merged into
a single pair and run from $(\W_{\cpl}, K_{\cpl})$ directly to
$(\W_{\cmi}, K_{\cmi})$.

\subsubsection*{Category VI}
As the crossing of the ELN distribution becomes so deep that
$\varPhi_0+\varPhi_z<0$ [but still with $G(-1)>0$], 
$\Delta$ becomes negative even at $\N=-1$. At this point, the real
branch has disappeared leaving only two pairs of complex-$K$ branches and
one pair of complex-$\W$ branches.

\section{Conclusions}
\label{sec:conclusions}

We have studied the critical points of the DR of the fast flavor
conversion of the neutrino medium. These critical points are also the
end points of the DR branches with complex frequencies and/or wave
numbers. Applying this theory
 to the neutrino medium with the axial symmetry, we demonstrated
how the DR branches and 
instabilities emerge and evolve as the ELN distribution is changed
continuously. We showed that, as one starts from an ELN distribution
with no unstable DR branches and varies it continuously, the branch
points of $K=\fK(\W)$ first appear as the turning points on the
real branch(es) before moving to the upper and lower complex planes of
$\W$. In this process, convective instabilities always appear first and then
may evolve into absolute instabilities as the ELN distribution changes.

We have explicitly shown that the instability associated with a
complex-$\W$ DR branch can still be non-convective even in the
presence of complex-$K$ 
branches. We have also shown that fast flavor conversions can occur
even when the ``zero mode'' is stable.
Our theory of the critical points of the DR function provides a way of
systematically studying the DRs 
and instabilities of
the neutrino oscillation wave. It also contributes to a good theoretical
foundation for 
the future studies of collective 
neutrino oscillations in core-collapse supernovae and binary neutrino
star mergers such as those in Refs.~\cite{Wu:2017qpc,Abbar:2018shq}.

\section*{Acknowledgments}
We thank S.~Abbar for the useful discussion.
We acknowledge the support by the US DOE
NP grant No.\ DE-SC0017803 at UNM.

\bibliography{dr}

\end{document}